\documentclass[%
 reprint,
 amsmath,amssymb,
 aps,
]{revtex4-2}
\usepackage{amsthm}
\usepackage{amsmath}
\usepackage{amsfonts}
\usepackage{amssymb}
\usepackage{graphicx}

\graphicspath{{./}{Single_q_reactor/}}
\usepackage[margin=0.5in]{geometry}

\usepackage{caption}
\usepackage{subcaption}
\usepackage{tikz-cd} 
\usepackage{xcolor}
\usepackage{physics}
\usepackage{ulem}

\begin{document}

\preprint{APS/123-QED}

\title{A spectral model of power-law decay in natural and engineered systems}

\author{Bal\'{a}zs S\'{a}ndor$^1$, M\'{a}rk Honti$^1$, and Henrique Santos Lima$^2$}
 \affiliation{ $^1$ Budapest University of Technology and Economics Department of Hydraulic and Water Resources Engineering, M\"{u}egyetem rakpart 3, Budapest 1111, Hungary \\ $^2$ Centro Brasileiro de Pesquisas Físicas, Rua Xavier Sigaud 150, Rio de Janeiro 22290-180, Brazil }

\thanks{A footnote to the article title}%

\date{\today}

\begin{abstract}
We present a first-principles spectral mechanism for the emergence of $q$-exponential dilution and power-law relaxation in non-ideal transport systems. By modeling an incompletely mixed reactor as a layered diffusion matrix with an absorbing boundary, we demonstrate that macroscopic power-law tails depend on the geometric interaction between the initial tracer placement and the domain's boundary configuration. For a one-dimensional system, an asymmetric, volumetrically distributed initial concentration profile projects onto the low-wavenumber eigenmodes, generating an emergent Gamma distribution of relaxation rates; at an infinitesimal boundary layer thickness ($\Delta z \to 0$), this profile yields the  $q$-exponential decay function exactly across the entire temporal domain with $q = 5/3$. Extended to $d$ dimensions under a highly localized, boundary-adjacent singular initial condition, the resulting scaling exponents and corresponding $q$ values depend explicitly on the spatial configuration of the absorbing boundaries. However, in the one-dimensional limit ($d=1$), these distinct initial states and boundary formulations intersect, rendering the $q=5/3$ exponent geometrically invariant. Our approach establishes a clear connection between linear diffusion transport and nonextensive statistical mechanics, showing how heavy-tailed transport can be derived from boundary geometry and spectral dimensionality.
\end{abstract}

\maketitle

\section*{Introduction}

Power-law relaxation is a characteristic feature of transport in highly heterogeneous systems. In fields ranging from non-ideal reactor engineering to transport in complex media, tracer concentrations systematically display algebraic tails of the form $C(t) \propto t^{-\beta}$ instead of the classical exponential decay expected from a single characteristic timescale \cite{Scher1975,Havlin1987,Bouchaud1990,Tsang1995,Metzler2000,Metzler2004,Chechkin2004,Sokolov2005,Berkowitz2006,Koren2007,Dentz2004,Haggerty1995,Hyman2019,DeSimone2023,Wang2023}. These persistent tails are macroscopic manifestations of multiscale exchange, geometric connectivity, and memory effects embedded within the transport domain. In statistical mechanics, a compact mathematical description of such non-exponential relaxation is provided by the  $q$-exponential function, which generalizes standard Boltzmann-Gibbs exponential decay and exhibits long-time algebraic scaling \cite{Tsallis1988,Tsallis1995,Plastino1995,Tsallis1999,Frank2004}. While this framework successfully reproduces anomalous transport profiles across various complex media, the entropic index $q$ is frequently introduced as a phenomenological fitting parameter \cite{Plastino1995,Borland1998,Bologna2000,Beck2001,Rangarajan2000,Niven2006,Beck2003}, leaving its mechanistic connection to the underlying physical transport operators incomplete.

This theoretical gap is clear when comparing the classical continuously stirred tank reactor (CSTR)—where ideal mixing yields a single relaxation mode and purely exponential dilution \cite{Rao1973,Nauman1983,Villermaux1986}—with structurally heterogeneous or incompletely mixed media. Real-world industrial reactors often operate in an incompletely mixed state, where stagnant regions, spatial bottlenecks, and multiscale mass transfer introduce a complex spectrum of decay rates that distort the expected residence time distribution \cite{Haggerty1995,Dentz2004,Berkowitz2006,Tsang1995,Hyman2019,DeSimone2023,Wang2023}. Rather than being governed by a single characteristic timescale, these systems possess intrinsic spectral heterogeneity that generates heavy-tailed tracer responses.

A central challenge lies in determining how a macroscopic $q$-exponential decay can arise from an explicit transport mechanism without assuming the distribution of relaxation rates \textit{a priori}. Under the paradigm of Beck-Cohen superstatistics, a $q$-exponential function can be formally constructed by averaging standard exponentials over a Gamma-distributed rate density \cite{Beck2001,Beck2003}. However, a first-principles derivation that maps a physically interpretable diffusion operator and a specific initial state directly to this emergent rate distribution has remained elusive. Establishing such a link is essential to determining whether $q$ is simply an empirical descriptor or a fundamental consequence of geometry, dimensionality, and spectral organization.

In this paper, we analyze a layered diffusion model with an absorbing top boundary, which serves as a mechanistic analogue for an incompletely mixed reactor \cite{Rao1973,Nauman1983,Villermaux1986}. We demonstrate that a spatially distributed, asymmetric initial tracer concentration profile projects onto the low-wavenumber eigenmodes of the diffusion operator \cite{Martelli2003,Carr2016} with modal weights scaling as $w(k) \propto k^2$. When combined with the standard diffusive dispersion relation $\omega(k) \propto -k^2$, this spectral structure leads analytically to a long-time dilution tail of $C_1(t) \sim t^{-3/2}$, yielding the entropic index $q = 5/3$ without secondary assumptions \cite{Tsallis1995,Plastino1995,Borland1998,Bologna2000,Tsallis1999,Egolf2018}. Crucially, by evaluating this analytical solution at an infinitesimal boundary layer thickness ($\Delta z \to 0$), the high-frequency exponential regularization terms vanish, proving that this asymmetric volume initialization generates the  $q$-exponential decay function exactly across the entire temporal domain. Furthermore, we extend the spectral analysis to $d$-dimensional configurations under a highly localized, boundary-adjacent singular initial condition, demonstrating that the resulting scaling exponents reflect the spatial geometry of the absorbing boundaries. We show that while these multi-dimensional boundary formulations yield distinct exponents for $d > 1$, they intersect seamlessly in the one-dimensional limit ($d = 1$), rendering the $q = 5/3$ exponent geometrically invariant. Our approach establishes a clear bridge between linear diffusion transport and nonextensive statistical mechanics \cite{Tsallis1988,Beck2001,Beck2003}, demonstrating that heavy-tailed transport and complex residence times can emerge solely from the interaction between an operator's spectral structure and boundary geometry.
\section*{Model and Methods}

\subsection*{Phenomenological dilution law}

We establish the macroscopic, phenomenological relaxation framework that the microscopic model must reproduce \cite{Tsallis1988,Tsallis1995,Plastino1995,Tsallis1999,Niven2006,Frank2004}. To capture non-ideal transport dynamics, the observable concentration $C(t)$ is assumed to satisfy the nonlinear relaxation equation 
\begin{equation}
\frac{\mathrm{d}C}{\mathrm{d}t}=-aC^{q}\,,
\end{equation}
where the entropic index $q$ parameterizes the structural departure from ideal first-order exponential decay. Under the initial condition $C(0)=(2-q)\lambda$ and with the scale parameter defined as $a=\lambda^{2-q}(2-q)^{1-q}$, the analytical solution is expressed via the generalized $q$-exponential function:
\begin{equation}
C(t)=(2-q)\lambda e_q^{-\lambda t}\,.
\end{equation}
Consequently, for $q>1$, the dilution curve accounts for systems governed by a continuous spectrum of active timescales rather than a single characteristic rate. In this  regime, the concentration profile exhibits a heavy algebraic tail that scales asymptotically as $C(t) \sim (\lambda t)^{-1/(q-1)}$.

\subsection*{Spectral and superstatistical formulation}

This  functional form maps directly onto a spectral representation. Within the framework of Beck-Cohen superstatistics \cite{Beck2001,Beck2003}, a macroscopic $q$-exponential decay emerges from a continuous superposition of standard first-order exponential relaxation processes,
\begin{equation}
C(t)=(2-q)\lambda\int_0^\infty \mathrm{d}\omega\, \tilde{w}(\omega)e^{-\omega t}=(2-q)\lambda e_q^{-\lambda t}\,,
\end{equation}
provided that the relaxation-rate density $\tilde{w}(\omega)$ follows a Gamma distribution. This equivalence motivates the core objective of this study: deriving such an effective distribution of relaxation rates from the first-principles mechanics of a linear transport operator, without invoking further statistical assumptions.

To address this problem, we formalize the internal reactor dynamics as a discrete, layered linear system. Consistent with eigenfunction treatments of layered diffusion media \cite{Martelli2003,Carr2016}, the evolution of the system is governed by the master equation 
\begin{equation}
\frac{\mathrm{d}\mathbf{C}}{\mathrm{d}t}=\mathbf{A}\mathbf{C}\,,\label{eq:master}
\end{equation}
where $\mathbf{C}(t) \in \mathbb{R}^n$ represents the state vector of individual layer concentrations and $\mathbf{A} \in \mathbb{R}^{n \times n}$ is the tridiagonal diffusion matrix incorporating localized boundary mass losses. The observable outflow signal is monitored via the concentration in the upper boundary layer, $C_1(t)$. Following diagonalization of the transport operator and taking the continuum limit ($n \to \infty$), the boundary concentration can be expanded as a spectral integral over wavenumber space:
\begin{equation}
C_1(t)\propto \int_0^\infty \mathrm{d}k\,w(k)e^{\omega(k)t}\,,
\end{equation}
where $w(k)$ denotes the structural modal weight and $\omega(k)<0$ is the eigenvalue dispersion relation associated with wavenumber $k$. As demonstrated below, the long-time power-law tail is entirely determined by the asymptotic structure of this spectral representation in the low-wavenumber ($k \to 0$) limit.

\section*{Results}
\subsection*{Homogeneous layered diffusion model}

We consider a discrete transport domain modeled as a layered network consisting of $n$ homogeneous intervals of width $\Delta z$, with a total characteristic length $L=n\Delta z$. The continuous-time evolution incorporates linear exchange coupling between adjacent layers, and a linear escape flux through the upper boundary layer is governed by Eq. \eqref{eq:master} , where $\mathbf{C}(t)$ is the state vector representing the concentration profile across the layers. The tridiagonal transport operator $\mathbf{A}$ representing homogeneous discrete diffusivity is given by 
\begin{equation}
\mathbf{A}=\tau
\begin{pmatrix}
-(1+\alpha)
& \alpha & 0 & 0 & \cdots & 0 \\
\alpha
& -2\alpha
& \alpha & 0 & \cdots & 0 \\
0
& \alpha
& -2\alpha
& \alpha & \cdots & 0 \\
\vdots & \ddots & \ddots & \ddots & \ddots & \vdots \\
0 & \cdots & 0 & \alpha
& -2\alpha
& \alpha\\
0 & \cdots & 0 & 0
& \alpha
& -\alpha
\end{pmatrix}\,.
\end{equation}
Here, $\tau$ represents the characteristic escape rate through the absorbing boundary, while $\alpha\tau$ defines the internal interlayer diffusion rate. The diagonal element $A_{11} = -\tau(1+\alpha)$ explicitly accounts for the localized mass loss at the upper boundary layer , whereas the remaining tridiagonal entries constitute a homogeneous discrete Laplacian operator with a zero-flux boundary condition at the bottom layer ($A_{nn} = -\alpha\tau$).

The eigenvalues of $\mathbf{A}$ are strictly negative and, for the transport regime where $\alpha>1/2$, occupy the bounded interval 
\begin{equation}
\omega\in[-4\alpha\tau\,,\,0] \,.
\end{equation}
The discrete dispersion relation mapping the wavenumber $k$ to the relaxation spectrum is derived as 
\begin{equation}
\omega(k)=-4\alpha\tau\sin^2\left(\frac{k}{2}\right)\,,\quad 0<k<\pi \,.
\end{equation}
By executing an eigenmode decomposition, the observable concentration at the boundary layer $C_1(t)$ is expanded as the spectral sum 
\begin{equation}
C_1(t)=\sum_{m=1}^n w_m e^{\omega_m t}\,,\label{eq:numerical}
\end{equation}
where $w_m$ denote the discrete modal weights.

The long-time asymptotic behavior of the system is governed by the relaxation modes situated near the spectral origin, corresponding to the low-wavenumber ($k \to 0$) limit. In this asymptotic regime, the dispersion relation simplifies to the standard diffusive scaling 
\begin{equation}
\omega(k)\sim -\alpha\tau k^2\,.
\end{equation}
For the class of distributed initial conditions derived below, the low-wavenumber modal weights scale quadratically as 
\begin{equation}
w(k)\sim k^2\,.
\end{equation}
Consequently, the discrete boundary flux scales asymptotically as 
\begin{equation}
C_1(t)\sim\sum_{m=1}^n k^2e^{-\alpha\tau k^2t}\,,\quad k=\frac{m\pi}{n}\,.
\end{equation}
Taking the continuum limit ($n \to \infty$) converts the sum into a spectral integral over the semi-infinite wavenumber domain: 
\begin{equation}
\lim_{n\to\infty} C_1(t)\sim\int_0^\infty \mathrm{d}k\,k^2\,e^{-\alpha\tau k^2t}=\frac{\sqrt{\pi}}{4(\alpha\tau t)^{\frac{3}{2}}}\propto t^{-\frac{3}{2}}\,.
\label{eq18}
\end{equation}
Matching this power-law exponent with the asymptotic tail of a generalized $q$-exponential function establishes that the  entropic index is uniquely constrained to $q=5/3$.

\subsection*{Isotropic extension to arbitrary dimensions}

The spectral scaling framework generalizes directly to an isotropic, $d$-dimensional discrete spatial lattice. Let the system state be defined by the concentration $C_{\mathbf{j}}(t)$ at each discrete node $\mathbf{j} = (j_1, j_2, \dots, j_d)$, where each coordinate index $j_\mu \in \mathbb{N}^+$ denotes the position along the $\mu$-th spatial dimension with a uniform lattice spacing $h$ that is identical for all dimensions. For a concrete three-dimensional system, this configuration simplifies to the standard notation $C_{i,j,k}(t)$. The master equation \eqref{eq:master} generalizes to the $d$-dimensional bulk lattice as 
\begin{equation}
\frac{\mathrm{d}}{\mathrm{d}t}C_{\mathbf{i}}=-\alpha \tau \sum_{\langle \mathbf{j} \rangle}(C_{\mathbf{i}}-C_{\mathbf{j}})\,,\label{eq:masterddim}
\end{equation}
where $\langle \mathbf{j}\rangle$ denotes the summation over all nearest-neighbor nodes.

The transport domain is bounded by absorbing layers at $j_\mu = 0$, enforcing homogeneous Dirichlet boundary conditions. Consequently, the separable multidimensional eigenmodes of the discrete transport matrix take the form
\begin{equation}
v_{\mathbf{k}}(\mathbf{j}) \propto \prod_{\mu=1}^d \sin(k_\mu j_\mu h)\,.
\end{equation}

We consider a singular initial concentration profile localized entirely at the corner node adjacent to the absorbing boundaries. Mathematically, this is expressed as the product of discrete Kronecker deltas:
\begin{equation}
C_{\mathbf{j}}(0) \propto \prod_{\mu=1}^d \delta_{j_\mu, 1}\,.
\end{equation}
The projection coefficients $a_{\mathbf{k}}$ of this initial state onto the eigenmodes are obtained by summing over the discrete spatial lattice:
\begin{equation}
a_{\mathbf{k}} = \sum_{j_1=1}^\infty \dots \sum_{j_d=1}^\infty C_{\mathbf{j}}(0) v_{\mathbf{k}}(\mathbf{j}) \propto \prod_{\mu=1}^d \sin(k_\mu h)\,.
\end{equation}
The observable temporal signal at the macroscopic boundary, $C_{\mathbf{1}}(t)$, corresponds to the concentration evaluated at the node $\mathbf{j} = (1, 1, \dots, 1)$. The discrete directional modal weights $w(\mathbf{k})$ for this node are given by $w(\mathbf{k}) = a_{\mathbf{k}} v_{\mathbf{k}}(\mathbf{1})$, which evaluates to:
\begin{equation}
w(\mathbf{k}) \propto \prod_{\mu=1}^d \sin^2(k_\mu h)\,.
\end{equation}
In the long-time asymptotic regime, the dynamics are governed by the low-wavenumber limit ($k_\mu h \ll 1$). Taylor expanding the modal weights yields the directional scaling density:
\begin{equation}
w(\mathbf{k}) \propto \prod_{\mu=1}^d k_\mu^2\,.
\end{equation}
Simultaneously, the multidimensional discrete dispersion relation simplifies to the standard continuous diffusive structure $\omega(\mathbf{k}) \sim -\alpha\tau \sum_{\mu=1}^d k_\mu^2 = -\alpha\tau k^2$.

Taking the continuum limit of the discrete spectral sum transforms it into a multidimensional integral over the wavenumber space. To evaluate the total boundary signal, we turn to hyperspherical coordinates, introducing a radial volume element $k^{d-1}\mathrm{d}k$. Integrating the directional modal weight density over the angular coordinates of the $d$-dimensional hypersphere isolates the effective radial modal weight. Because the product $\prod_{\mu=1}^d k_\mu^2$ is a homogeneous polynomial of degree $2d$, its angular integral evaluates, by isotropic symmetry, to a term strictly proportional to $k^{2d}$.

\begin{figure}[htb]
    \centering
    \includegraphics[width=1.0\linewidth]{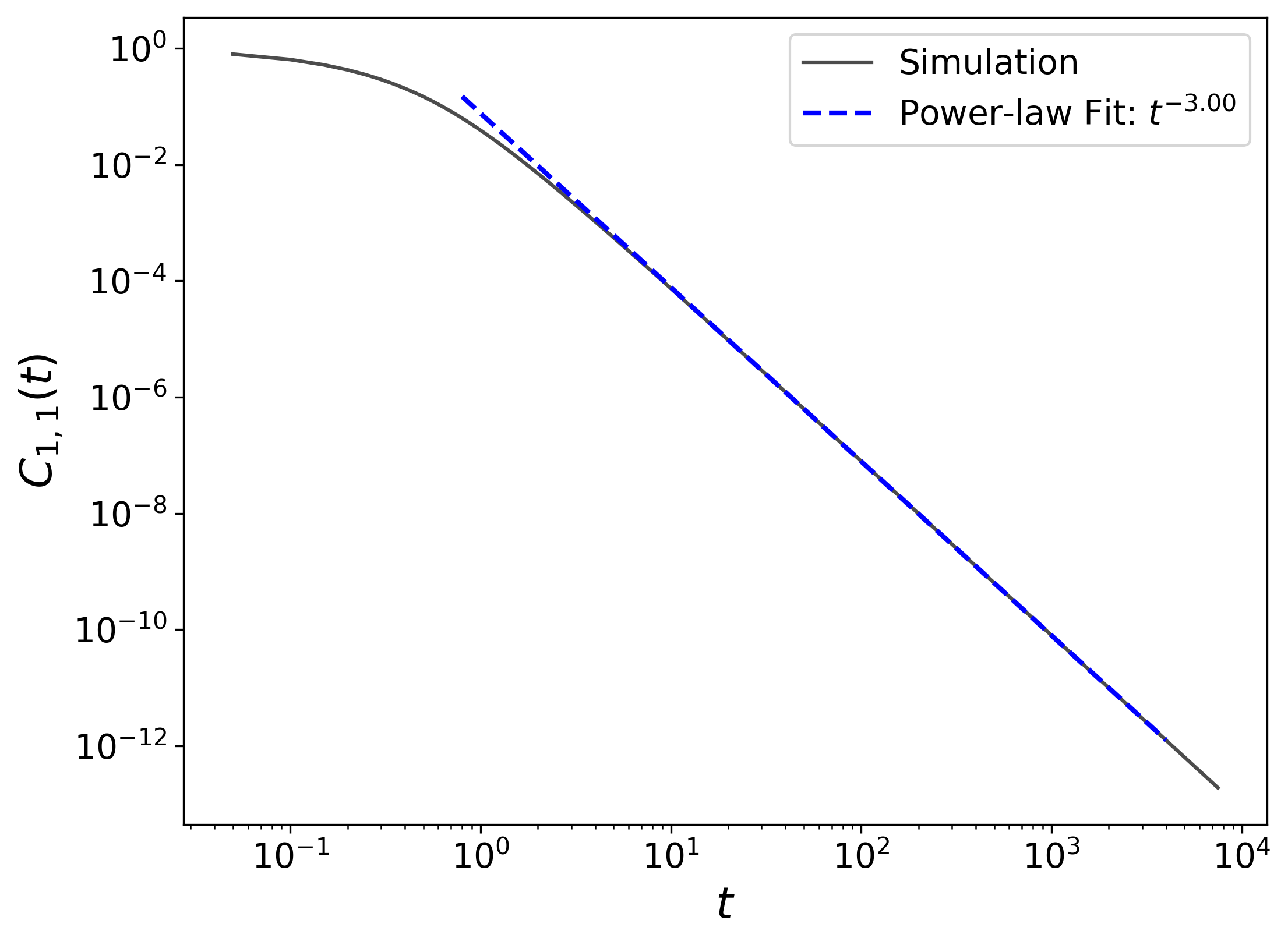}
    \caption{Boundary concentration $C_{1,1}(t)$ as a function of time $t$ for the two-dimensional transport model (Eq. \eqref{eq:masterddim} for $d=2$). The numerical simulation of the discrete layered diffusion matrix under absorbing Dirichlet boundary conditions on a  $512 \times 512$ grid (solid black line) is consistent with the  analytical asymptotic scaling law $t^{-3}$ (dashed blue line).}
    \label{fig:placeholder}
\end{figure}

The resulting integration over the radial isotropic wavenumber space determines the exact temporal scaling of the boundary signal:
\begin{equation}
C_{\mathbf{1}}(t)\sim \int_0^\infty \mathrm{d}k\,k^{d-1}\,k^{2d}\,e^{-\alpha\tau k^2t}\propto t^{-3d/2}\,.\label{eq:asymptotic}
\end{equation}
Equating this result to the characteristic algebraic tail of the generalized $q$-exponential function, $t^{-1/(q-1)}$, establishes the algebraic system
\begin{equation}
\frac{1}{q-1}=\frac{3d}{2}\,,\qquad q=1+\frac{2}{3d}\,.\label{eq:qd}
\end{equation}
This scaling relation dictates specific entropic thresholds for physical dimensions, yielding $q=5/3$, $4/3$, and $11/9$ for $d=1,2,3$, respectively. This multi-dimensional scaling behavior is verified numerically in FIG. \ref{fig:placeholder} for a two-dimensional system ($d=2$), demonstrating a robust agreement between the full discrete matrix simulation on a fine grid and the predicted $t^{-3}$ asymptotic algebraic decay regime. 
 
In the thermodynamic mean-field limit ($d\to\infty$), localized spectral fluctuations vanish, and the system asymptotically recovers standard exponential relaxation ($q\to1$).

For $d=1, 2, 3$, the system possesses 2, 4, and 6 nearest neighbors, respectively. As $d \to \infty$, the local coordination number diverges ($2d \to \infty$), meaning that every layer becomes strongly connected to every other layer. In the thermodynamic mean-field limit ($d\to\infty$), localized spectral fluctuations vanish, and the system's behavior can be described by mapping the lattice onto an effective complete graph where every node couples directly to the global average. This infinite-range mean-field proxy is represented as follows:  

\begin{equation}
\frac{\mathrm{d}}{\mathrm{d}t}C_i=-\frac{\alpha \tau}{N}\sum_{j=1}^N (C_i-C_j).
\end{equation}
Notice that 
\begin{equation}
    \frac1{N}\sum_j C_j=\langle C \rangle\,,
\end{equation}
and $\frac{\mathrm{d}}{\mathrm{d}t}\langle C\rangle=0$ which implies that $\langle C \rangle$ is a constant of motion. Therefore, knowing that $\sum_j C_i=N C_i$

\begin{equation}
    \frac{\mathrm{d}}{\mathrm{d}t}C_i=-\frac{1}{t_r}(C_i-\langle C \rangle)\,,
\end{equation}
which is a CSTR-like equation. The characteristic time  $t_r\equiv 1/(\alpha \tau)$ is defined as the relaxation time of this process.

Let us emphasize that the application of homogeneous Dirichlet boundary conditions yields a mathematically consistent class of asymptotic dilution solutions, as described by Eq. \eqref{eq:asymptotic} for arbitrary dimensions. However, the exact scaling behavior remains structurally coupled to the geometric manifold of the absorbing boundary. If the absorbing Dirichlet condition is restricted to a single $(d-1)$-dimensional top-layer hyperplane by imposing $C_{0,j_2,\dots,j_d}(t) = 0$ while enforcing Neumann (reflective) boundary conditions along the remaining $d-1$ axes, the boundary flux scales as $C_1(t) \sim t^{-(d+2)/2}$. This configuration yields the modified  scaling thresholds $1/(q-1) = 3/2, 2,$ and $5/2$ for dimensions $d=1,2,3$, respectively. Mechanistically, the introduction of reflective boundaries alters the separable multidimensional eigenmodes, generating cosine functions along the Neumann axes and a single sine function along the absorbing Dirichlet axis. In the low-wavenumber limit ($k_\mu h \to 0$), the cosine components converge to unity, eliminating their directional contribution to the scaling density. Consequently, the directional modal weights no longer scale as $w(\mathbf{k}) \propto k^{2d}$ but collapse to a single parabolic contribution $w(\mathbf{k}) \propto k^2$ dictated entirely by the absorbing axis. Integrating this localized weight density over the isotropic hyperspherical volume element $k^{d-1}\mathrm{d}k$ directly recovers the modified temporal power-law decay.

Both geometric formulations intersect seamlessly at $d=1$, where a localized boundary corner and a boundary hyperplane are topologically equivalent.

 Similar classification for first passage time distributions for $d$ dimensional Gaussian process (even functions with $w(k)\sim1$ wavenumbers) reads as \cite{Redner2001}
\begin{equation}
C_1(t)\sim
\begin{cases}
t^{\frac{d}{2}-2}\,,& d<2\\
\frac{1}{t\ln^2t}\,,& d=2\\
t^{-\frac{d}{2}}\,,& d>2
\end{cases}\,.
\end{equation}

\subsection*{Superstatistical interpretation of the spectral weights}

\begin{figure}[htb]
\centering
\includegraphics[width=0.8\linewidth]{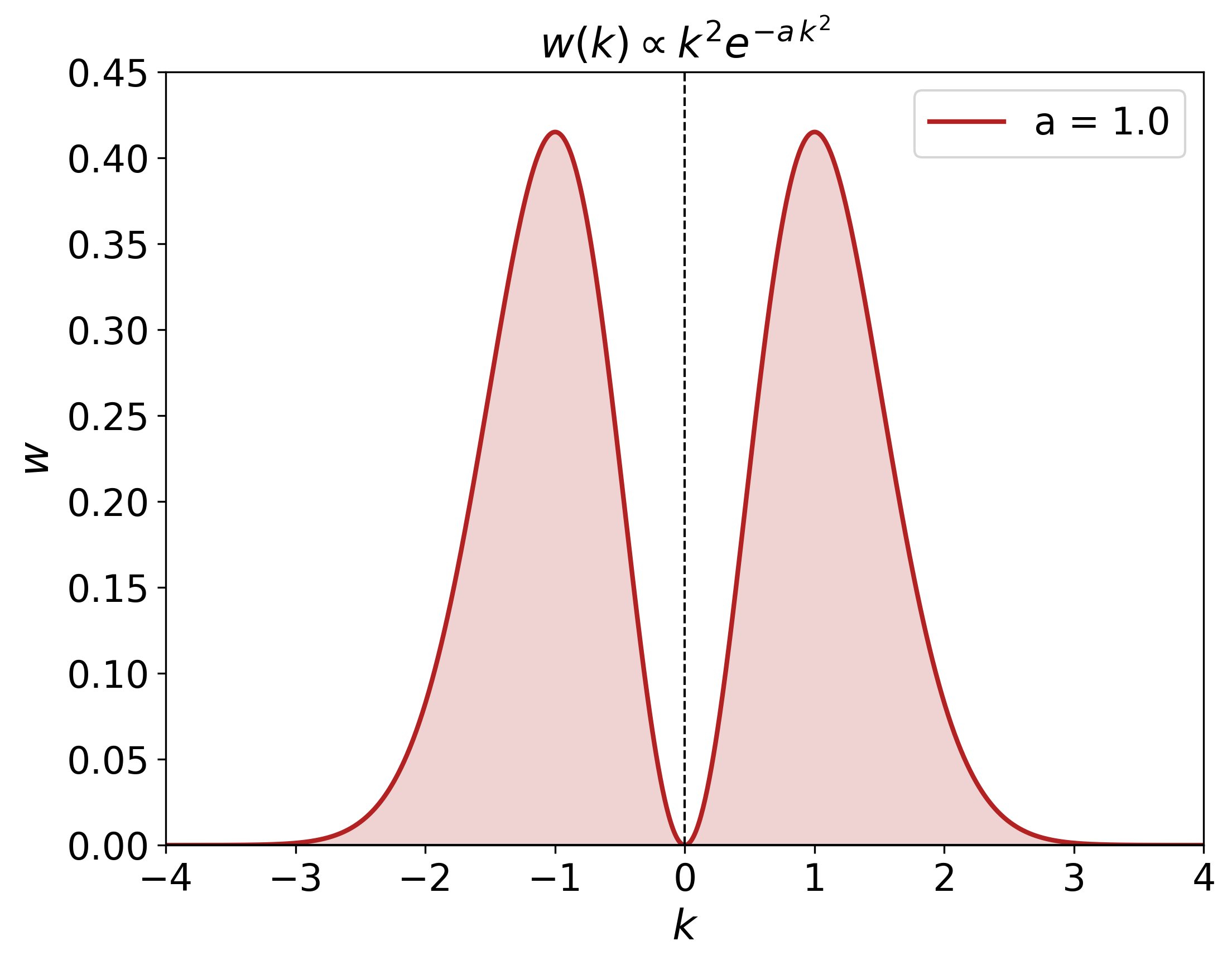}
\caption{Modal weight distribution $w(k)$ derived from the emergent Gamma rate spectrum in Eq. \eqref{wgamma}.}
\label{fig:modalfull}
\end{figure}

The spectral expansion of the boundary concentration maps directly onto an integral Laplace transform over a continuous domain of positive relaxation rates. Substituting the variable transformation $\omega=\alpha\tau k^2$ into the low-wavenumber spectral integral,
\begin{equation}
    C_1(t)\sim \int_0^\infty \mathrm{d}k\, w(k) e^{-\alpha \tau k^2t}\,,
\end{equation}
yields the superstatistical representation
\begin{equation}
    C_1(t)\propto \int_0^\infty \mathrm{d}\omega\,\tilde{w}(\omega) e^{-\omega t}\,,
\end{equation}
where the effective relaxation-rate density $\tilde{w}(\omega)$ is related to the structural modal weights via the Jacobian transformation
\begin{equation}
\tilde{w}(\omega)=\frac{w(k(\omega))}{2\alpha\tau k(\omega)}\,.
\end{equation}
Enforcing the condition that $\tilde{w}(\omega)$ satisfies a standard Gamma distribution,
\begin{equation}
    \tilde{w}(\omega)=\frac{a^n \omega^{n-1}}{\Gamma(n)} e^{-a \omega}\,,
\end{equation}
the integration evaluates exactly to
\begin{equation}
    C_1(t) \propto \frac{a^n}{\Gamma(n)}\int_0^\infty \mathrm{d}\omega\, \omega^{n-1} e^{-(a+t) \omega} =[1+t/a]^{-n}\,.
\end{equation}
Imposing the parameter mapping $a=n/\lambda$ and $n=1/(q-1)$, this structural form matches the  $q$-exponential distribution:
\begin{equation}
    C_1(t) \propto [1+(q-1)\lambda t]^{-1/(q-1)}\,.
\end{equation}
For the specific index $q=5/3$, the asymptotic long-time decay reduces to $C_1(t)\sim A t^{-3/2}$, where $A>0$. Transforming the rate density back to the wavenumber domain yields the explicit distribution (FIG. \ref{fig:modalfull}):
\begin{equation}
    w(k)=\frac{4}{\sqrt{\pi}}\left(\frac{3\alpha\tau}{2\lambda}\right)^{3/2} k^2 e^{-\frac{3\alpha\tau k^2}{2\lambda}}\,.
    \label{wgamma}
\end{equation}
In the low-wavenumber limit ($k \to 0$), this analytical density reduces to the scaling $w(k)\propto k^2$, verifying the origin of the $t^{-3/2}$ power-law tail.

The entropic index $q$ can be formally interpreted as a normalized statistical measure of spectral heterogeneity. For the derived Gamma rate distribution, the variance of the underlying wavenumber space satisfies
\begin{equation}
    q-1=\frac{1}{n}=\frac{\langle k^4 \rangle-\langle k^2 \rangle^2}{\langle k^2 \rangle^2}\,.
\end{equation}
When the fluctuations in $k^2$ approach zero, the index $q \to 1$, mapping the system back to a single-rate first-order exponential relaxation regime (see Fig. \ref{fig:modalfull}).

\section*{Condition for the emergence of a $q$-exponential law}

We derive the specific class of boundary conditions and initial states required to generate the quadratic low-wavenumber modal scaling. Rather than initializing the system with a singular point-source impulse ($\delta_{j,1}$) at the boundary layer---which introduces unphysical high-frequency transients---we consider a macroscopically regularized, volume-distributed initial profile:
\begin{equation}
C_j(0) \propto j\Delta z e^{-\frac{\lambda}{6\alpha\tau}(j\Delta z)^2}.
\end{equation}
This asymmetric distribution starts at zero at the open boundary, scales linearly ($j\Delta z$) in the near-field, and is bounded by a Gaussian exponential cutoff in the far-field. This distributed configuration smoothly loads the low-wavenumber modes of the reactor from the onset. The time-dependent solution to the linear system is expanded in the eigenbase of the transport operator $\mathbf{A}$ as
\begin{equation}
\mathbf{C}(t)=\sum_{k} a_k \mathbf{v}_k e^{\omega_k t},
\end{equation}
where $a_k$ denote the projection coefficients and $\mathbf{v}_k$ represent the orthonormal eigenvectors.

The absorbing boundary condition enforced at the upper surface implies the Dirichlet constraint $v_{0,k}=0$. Consequently, in the low-wavenumber regime, the spatial eigenvectors take the form
\begin{equation}
v_{j,k}\propto \sin(kj\Delta z)\,.
\end{equation}
The discrete modal contribution to the observable boundary layer is determined by the projection
\begin{equation}
    w_k\propto a_k v_{1,k}\,.
\end{equation}
Utilizing internal orthonormality, the projection coefficients $a_k$ are evaluated via the inner product
\begin{equation}
\begin{aligned}
a_k = \mathbf{C}(0)\cdot \mathbf{v}_k 
&\propto \sum_{j=1}^n C_j(0) \sin(k j \Delta z) \\
&\propto \sum_{j=1}^n j\Delta z \, e^{-\frac{\lambda}{6\alpha\tau} (j\Delta z)^2} \sin(k j \Delta z)\,.
\end{aligned}
\end{equation}
Evaluating this sum in the continuum limit yields the integral expression
\begin{equation}
    a_k\sim \int_0^\infty \mathrm{d}z \,z e^{-\frac{\lambda}{6\alpha\tau} z^2} \sin(k z)\propto k e^{-\frac{3\alpha\tau k^2}{2\lambda}}\,.
\end{equation}
Since the boundary eigenvector components scale as $v_{1,k}\propto \sin(k\Delta z)\simeq k\Delta z$ for $k\Delta z\ll1$, the structural modal weights satisfy
\begin{equation}
    w_k\propto k^2 e^{-\frac{3\alpha\tau k^2}{2\lambda}}\,.
    \label{mwnew}
\end{equation}
This confirms that the distributed initial profile generates the $w(k)\propto k^2$ scaling required in Eq. \eqref{eq18}.

\begin{figure}[htb]
\centering
\includegraphics[width=1\linewidth]{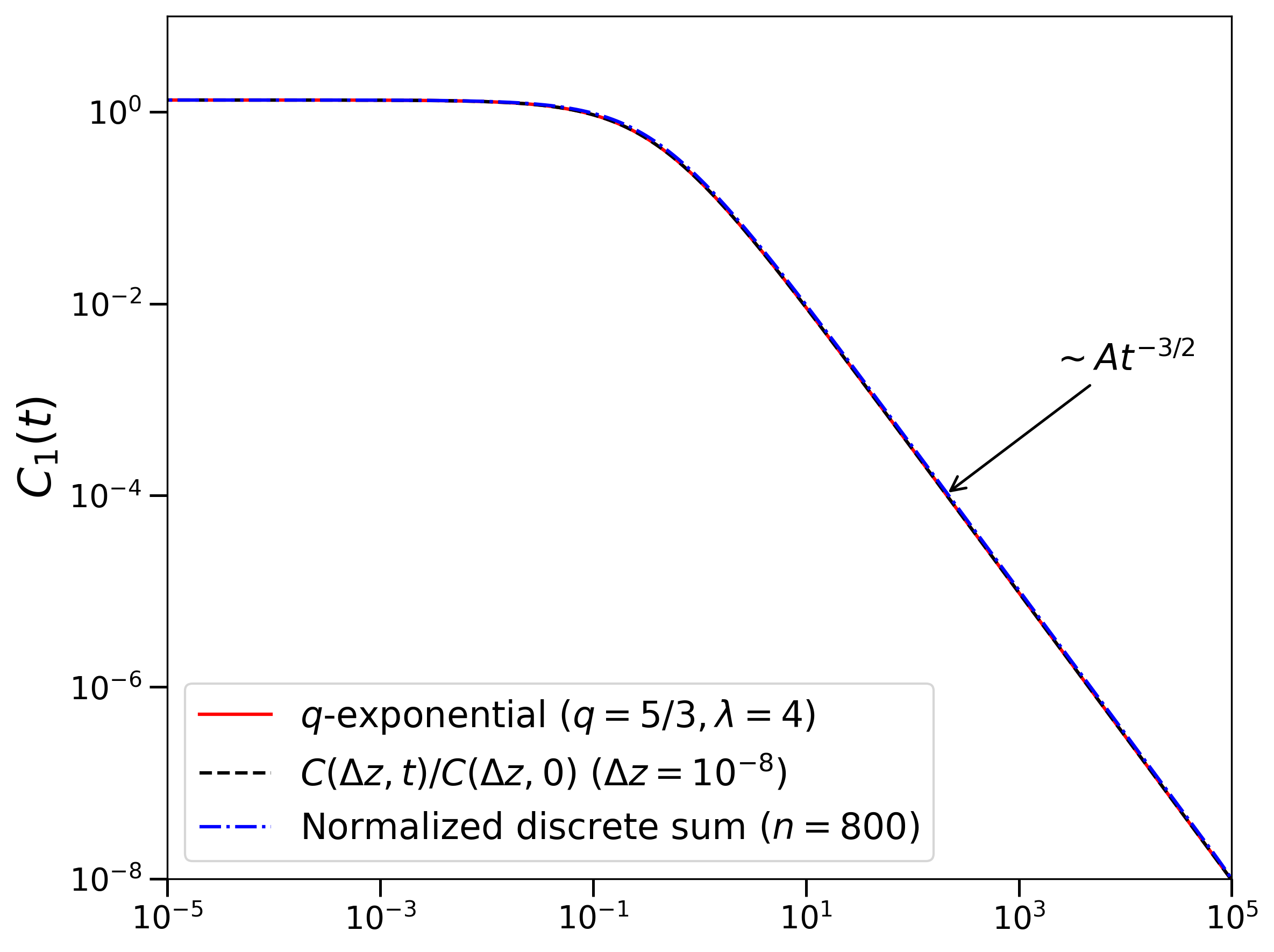}\\
\includegraphics[width=1.0\linewidth]{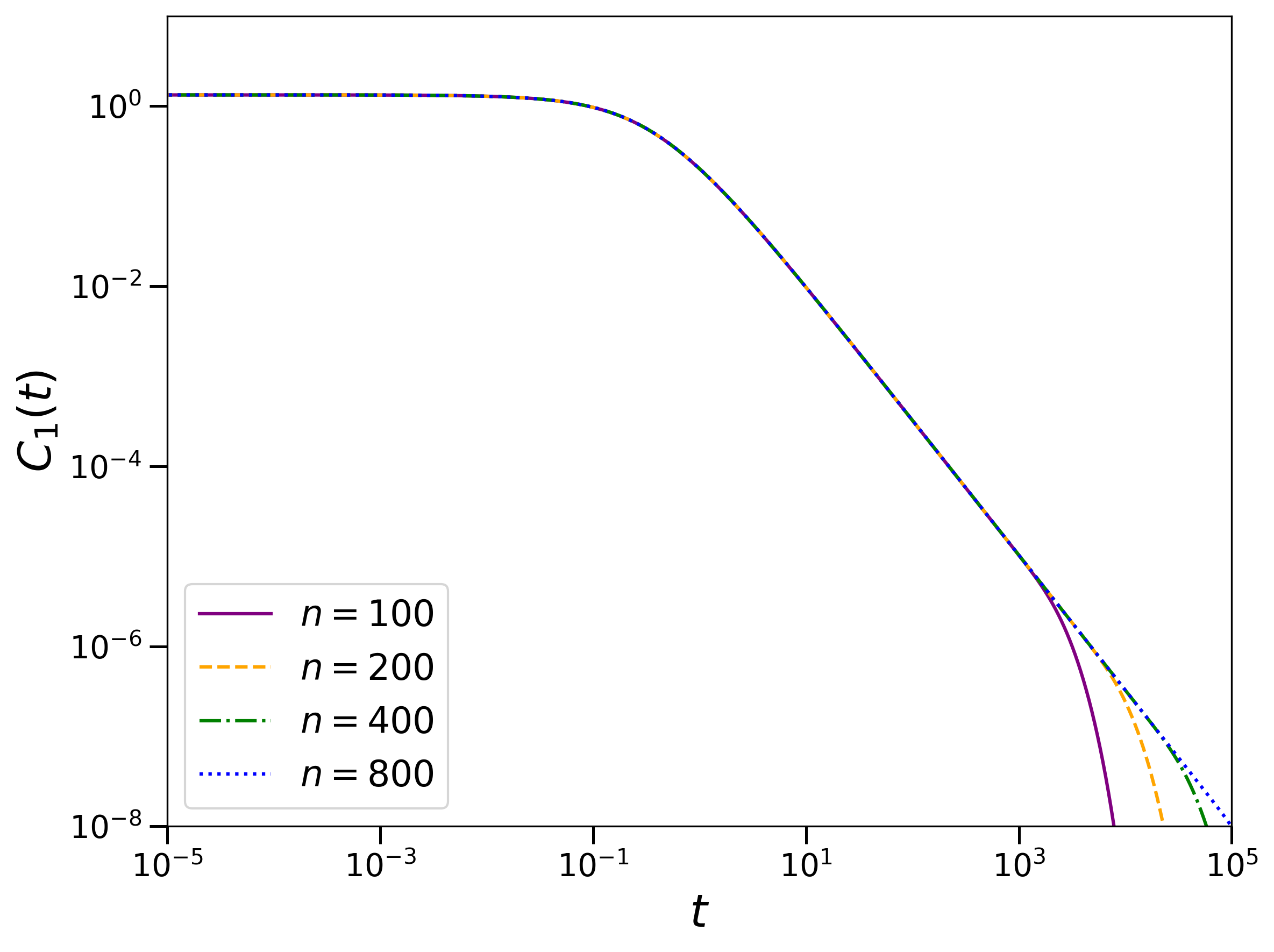}
\caption{(Top) Boundary concentration $C_1(t)$ as a function of time $t$ for the normalized $q$-exponential function (red solid curve), the continuous analytical solution in Eq. \eqref{eqshifted} (black dashed curve), and the discrete spectral sum computed for $n=800$ layers via Eq. \eqref{eq:numerical} and Eq. \eqref{mwnew} (blue dash-dotted curve). (Bottom) Discrete spectral sum for $n=100$ (purple), $n=200$ (orange), $n=400$ (green), and $n=800$ (blue).}
\label{figcomp}
\end{figure}

The corresponding continuous space-time concentration profile is evaluated via the inverse transform:
\begin{equation}
    C(z,t) \sim C_0 \int_0^\infty \mathrm{d}k\, k\, e^{-\frac{3\alpha\tau k^2}{2\lambda}} \sin(k z) e^{-\alpha \tau k^2 t}\,.
\end{equation}
Defining the time-dependent scale factor $A(t)=\frac{3\alpha\tau}{2\lambda}+\alpha\tau t$, explicit integration yields the analytical solution
\begin{equation}
    C(z,t)\propto \frac{z}{A(t)^{3/2}} e^{-z^2/(4 A(t))}\,,
    \label{eqshifted}
\end{equation}
where $A(t)^{-3/2}\propto e_{5/3}^{-\lambda t}$. Evaluating this profile at the boundary layer $z=\Delta z$ under the long-time asymptotic condition $(\Delta z)^2/t\ll1$, the expression reduces to the $t^{-3/2}$ scaling law, confirming the emergence of the $q=5/3$ tail (FIG. \ref{figcomp}). 

As demonstrated in the top panel of FIG. \ref{figcomp}, the discrete spectral sum computed for $n=800$ layers exhibits a robust agreement with both the continuous analytical field solution and the ideal  $q$-exponential decay function, validating the regularization mechanism across all temporal scales. The bottom panel of FIG. \ref{figcomp} illustrates the corresponding finite-size truncation effects inherent to the discrete formulation; as the layer cardinality $n$ increases, the algebraic power-law scaling regime systematically extends to longer temporal domains before transitioning into an exponential cutoff dictated by the bounded nature of the discrete transport matrix spectrum.

For comparison, a highly localized singular initial condition $C_j(0)=\delta_{j,1}$ produces the standard Green's function solution:
\begin{equation}
    C(z,t)\propto \frac{z}{t^{3/2}} e^{-z^2/(4 \alpha \tau t)}\,.
\end{equation}
While this localized profile converges to the same asymptotic long-time form $C(\Delta z,t)\approx B t^{-3/2}$ ($B>0$), the Gaussian-shifted initial profile operates as a physically regularized variant that preserves this scaling behavior across all temporal regimes.

\section*{Continuum interpretation}

The discrete matrix formulation possesses an exact continuum counterpart. We consider the one-dimensional diffusion equation
\begin{equation}
    \partial_t C=D\partial_z^2 C\,,
\end{equation}
defined on the semi-infinite domain $z \in [0, \infty)$, where $D = \alpha \tau (\Delta z)^2$ is the continuous diffusion coefficient, subject to the homogeneous Dirichlet boundary condition $C(0,t)=0$. Extending the solution space to the negative domain as an odd function of $z$ establishes the Fourier sine transform as the appropriate spectral operator. In wavenumber space, the decoupling of the differential operator yields
\begin{equation}
    \tilde{C}(k,t)=\tilde{C}(k,0)e^{-Dk^2 t}\,.
\end{equation}
The continuous inverse transformation back to real space is given by
\begin{equation}
    C(z,t)=\int_0^{\infty} \mathrm{d}k\,\tilde{C}(k,0)e^{-Dk^2 t}\sin(kz)\,.
\end{equation}
Imposing the continuous initial concentration profile
\begin{equation}
C(z,0)\propto z e^{-\frac{\lambda}{6\alpha\tau}z^2}\,,
\end{equation}
the forward Fourier sine transform evaluates to
\begin{equation}
    \tilde{C}(k,0)=\int_0^\infty \mathrm{d}z\, C(z,0)\sin(kz)\propto k e^{-\frac{3\alpha\tau k^2}{2\lambda}}\,.
\end{equation}
Direct substitution into the inverse transform reproduces the analytical structure of Eq. \eqref{eqshifted} under the matching continuous scaling. 

The continuum formulation demonstrates that the effective superstatistical rate distribution is an emergent property generated by the projection of the initial state onto the spectral eigenmodes of the diffusion operator. This unifies the discrete matrix transport model, the continuous parabolic field description, and the macroscopic emergence of the $q$-exponential dilution law.

\section*{Final remarks}

We have presented a first-principles spectral mechanism for the emergence of $q$-exponential dilution and power-law relaxation within a linear, layered transport operator incorporating boundary losses. The foundational result for one-dimensional systems is that a volumetrically distributed, asymmetric initial tracer concentration profile projects onto the low-wavenumber eigenmodes of the diffusion operator with modal weights scaling quadratically as $w(k) \propto k^2$. In conjunction with the parabolic diffusive dispersion relation $\omega(k) \propto -k^2$, this spectral structure leads analytically to a long-time dilution tail scaling as $t^{-3/2}$, which maps uniquely to the  entropic index $q = 5/3$. Crucially, in the continuum limit where the boundary layer thickness vanishes ($\Delta z \to 0$), the high-frequency exponential regularization terms disappear, establishing that this asymmetric volume initialization recovers the  $q$-exponential function exactly across the entire temporal domain.

This derivation provides an explicit, operator-theoretic foundation for nonextensive relaxation parameters that have conventionally been introduced on phenomenological grounds. Within this framework, the entropic index $q$ is not an empirical descriptor adjusted to fit observed non-exponential dilution curves; rather, it emerges directly from the small-eigenvalue structure of the transport matrix and the boundary projection of the initial state. The resulting Gamma distribution of relaxation rates is derived from first principles, establishing a direct mathematical bridge between linear diffusion processes, superstatistics, and macroscopic non-exponential tracer dynamics.

The multi-dimensional isotropic extension demonstrates that when evaluating a highly localized, boundary-adjacent singular initialization, the resulting scaling exponents systemically encode the boundary geometry and effective spatial dimensionality of the transport domain via the relation $q = 1 + 2/(3d)$. This relation recovers the distinct thresholds of $q = 5/3, 4/3,$ and $11/9$ for physical dimensions $d = 1, 2, 3$, and converges smoothly to the exponential mean-field limit ($q \to 1$) as $d \to \infty$. While the higher-dimensional scaling exponents depend explicitly on the spatial configuration of the absorbing boundaries—reflecting different structural layouts in engineering applications—they intersect seamlessly in the one-dimensional limit, rendering the $q = 5/3$ exponent geometrically invariant under this class of linear transport.

Although our structural model is parsimonious, it successfully captures the fundamental dynamics governing anomalous tracer tailing in incompletely mixed chemical or process reactors. These findings indicate that heavy-tailed tracer responses and non-exponential residence time distributions can be fundamentally understood through the spectral organization of interconnected diffusion fields.

A clear limitation of the current formulation is its reliance on a spatially homogeneous diffusion matrix, which leaves the direct connection between partially mixed dilution and non-integer effective dimensions open for further study. Characterizing anomalous relaxation across fractal or disordered configurations would provide a unified mathematical framework capable of connecting highly irregular structural networks found in complex natural and engineered systems. A particularly relevant extension for process engineering lies in analyzing a series or cascade of non-ideal reactors, which is a structural configuration highly representative of realistic chemical processing and wastewater treatment systems \cite{Stantec2022, Hurtado2015}. Computing flood waves in streams as cascade of reservoirs is also a potential application field, where analytical approach is only available for linearly dependent state-space variables using exponential and Gamma distributions \cite{Szollosi1982, Szilagyi2010}. Future investigations will focus on integrating these analytical scaling laws with multi-stage network models and testing them against empirical breakthrough curve data across alternative non-Dirichlet boundary configurations.

\section*{acknowledgments}
We acknowledge fruitful discussions with M. A. Pires as well as partial financial support from Conselho Nacional de Desenvolvimento Científico e Tecnológico
 (CNPq), Fundação Carlos Chagas Filho de Amparo à Pesquisa do Estado do Rio de Janeiro (FAPERJ), and Coordenação de Aperfeiçoamento de Pessoal de Nível Superior (CAPES) (Brazilian agencies). This article was produced with the support of the Erasmus+ project 2024-1-HU01-KA131-HED-000209516.


\begin{thebibliography}{99}

\bibitem{Scher1975}
H. Scher and E. W. Montroll.
\newblock \textit{Anomalous transit-time dispersion in amorphous solids.}
\newblock Physical Review B \textbf{12}(6), 2455--2477 (1975).

\bibitem{Havlin1987}
S. Havlin and D. Ben-Avraham.
\newblock \textit{Diffusion in disordered media.}
\newblock Advances in Physics \textbf{36}(6), 695--798 (1987).

\bibitem{Bouchaud1990}
J.-P. Bouchaud and A. Georges.
\newblock \textit{Anomalous diffusion in disordered media: Statistical mechanisms, models and physical applications.}
\newblock Physics Reports \textbf{195}(4--5), 127--293 (1990).

\bibitem{Tsang1995}
Y. W. Tsang.
\newblock \textit{Study of alternative tracer tests in characterizing transport in fractured rocks.}
\newblock Geophysical Research Letters \textbf{22}(11), 1421--1424 (1995).

\bibitem{Metzler2000}
R. Metzler and J. Klafter.
\newblock \textit{The random walk's guide to anomalous diffusion: a fractional dynamics approach.}
\newblock Physics Reports \textbf{339}(1), 1--77 (2000).

\bibitem{Metzler2004}
R. Metzler and J. Klafter.
\newblock \textit{The restaurant at the end of the random walk: Recent developments in the description of anomalous transport by fractional dynamics.}
\newblock Journal of Physics A: Mathematical and General \textbf{37}(31), R161--R208 (2004).

\bibitem{Chechkin2004}
A. V. Chechkin, V. Yu. Gonchar, J. Klafter, and R. Metzler.
\newblock \textit{Fundamentals of L\'evy flight processes.}
\newblock Advances in Chemical Physics \textbf{133}, 439--496 (2004).

\bibitem{Sokolov2005}
I. M. Sokolov and J. Klafter.
\newblock \textit{From diffusion to anomalous diffusion: A century after Einstein.}
\newblock In Anomalous Diffusion: From Basics to Applications, Springer (2005).

\bibitem{Berkowitz2006}
B. Berkowitz, A. Cortis, M. Dentz, and H. Scher.
\newblock \textit{Modeling non-Fickian transport in geological formations as a continuous time random walk.}
\newblock Reviews of Geophysics \textbf{44}(2), RG2003 (2006).

\bibitem{Koren2007}
T. Koren, M. A. Lomholt, A. V. Chechkin, J. Klafter, and R. Metzler.
\newblock \textit{First passage times of L\'evy flights coexisting with subdiffusion.}
\newblock Physical Review E \textbf{76}(3), 031129 (2007).

\bibitem{Dentz2004}
M. Dentz, A. Cortis, H. Scher, and B. Berkowitz.
\newblock \textit{Time behavior of solute transport in heterogeneous media: Transition from non-Fickian to Fickian dispersion.}
\newblock Physical Review E \textbf{70}(1), 010101(R) (2004).

\bibitem{Haggerty1995}
R. Haggerty and S. M. Gorelick.
\newblock \textit{Multiple-rate mass transfer for modeling diffusion and surface reactions in media with pore-scale heterogeneity.}
\newblock Water Resources Research \textbf{31}(10), 2383--2400 (1995).

\bibitem{Hyman2019}
J. D. Hyman, H. Rajaram, S. Srinivasan, N. Makedonska, S. Karra, H. Viswanathan, and G. Srinivasan.
\newblock \textit{Matrix diffusion in fractured media: New insights into power law scaling of breakthrough curves.}
\newblock Geophysical Research Letters \textbf{46}(23), 13785--13795 (2019).

\bibitem{DeSimone2023}
S. De Simone, O. Bour, and P. Davy.
\newblock \textit{Impact of matrix diffusion on heat transport through heterogeneous fractured aquifers.}
\newblock Water Resources Research \textbf{59}(2) (2023).

\bibitem{Wang2023}
L. Wang, S. Yoon, L. Zheng, T. Wang, X. Chen, and P. K. Kang.
\newblock \textit{Flux exchange between fracture and matrix dictates late-time tracer tailing.}
\newblock Journal of Hydrology \textbf{627}, 130480 (2023).

\bibitem{Tsallis1988}
C. Tsallis.
\newblock \textit{Possible generalization of Boltzmann-Gibbs statistics.}
\newblock Journal of Statistical Physics \textbf{52}(1--2), 479--487 (1988).

\bibitem{Tsallis1995}
C. Tsallis, S. V. F. Levy, A. M. C. Souza, and R. Maynard.
\newblock \textit{Statistical-mechanical foundation of the ubiquity of L\'evy distributions in nature.}
\newblock Physical Review Letters \textbf{75}(20), 3589--3593 (1995).

\bibitem{Plastino1995}
A. R. Plastino and A. Plastino.
\newblock \textit{Non-extensive statistical mechanics and the Fokker-Planck equation.}
\newblock Physica A: Statistical Mechanics and its Applications \textbf{222}(1--4), 347--354 (1995).

\bibitem{Tsallis1999}
C. Tsallis.
\newblock \textit{Nonextensive statistics: Theoretical, experimental and computational evidences and connections.}
\newblock Brazilian Journal of Physics \textbf{29}(1) (1999).

\bibitem{Frank2004}
T. D. Frank.
\newblock \textit{Nonlinear Fokker-Planck equations: Fundamentals and applications.}
\newblock Springer, Berlin (2004).

\bibitem{Borland1998}
L. Borland.
\newblock \textit{Microscopic dynamics of the nonlinear Fokker-Planck equation: A phenomenological model.}
\newblock Physical Review E \textbf{57}(6), 6634--6642 (1998).

\bibitem{Bologna2000}
M. Bologna, C. Tsallis, and P. Grigolini.
\newblock \textit{Anomalous diffusion associated with nonlinear fractional derivative Fokker-Planck-like equation: Exact time-dependent solutions.}
\newblock Physical Review E \textbf{62}(2), 2213--2218 (2000).

\bibitem{Beck2001}
C. Beck.
\newblock \textit{Dynamical foundations of nonextensive statistical mechanics.}
\newblock Physical Review Letters \textbf{87}(18), 180601 (2001).

\bibitem{Rangarajan2000}
G. Rangarajan and M. Ding.
\newblock \textit{First passage time distribution for anomalous diffusion.}
\newblock Physics Letters A \textbf{273}(5--6), 322--330 (2000).

\bibitem{Niven2006}
R. K. Niven.
\newblock \textit{Exact q-exponential solutions of the logistic equation and related reaction-diffusion models.}
\newblock Physical Review E \textbf{74}(3), 031111 (2006).

\bibitem{Beck2003}
C. Beck and E. G. D. Cohen.
\newblock \textit{Superstatistics.}
\newblock Physica A: Statistical Mechanics and its Applications \textbf{322}, 267--275 (2003).

\bibitem{Rao1973}
D. P. Rao and L. L. Edwards.
\newblock \textit{Mixing effects in stirred tank reactors: A comparison of models.}
\newblock Chemical Engineering Science \textbf{28}(5), 1179--1192 (1973).

\bibitem{Nauman1983}
E. B. Nauman and B. A. Buffham.
\newblock \textit{Mixing in continuous flow systems.}
\newblock John Wiley \& Sons, New York (1983).

\bibitem{Villermaux1986}
J. Villermaux.
\newblock \textit{The role of hydrodynamics in the design of chemical reactors.}
\newblock In Encyclopedia of Fluid Mechanics, Vol. 4, Gulf Publishing (1986).

\bibitem{Martelli2003}
F. Martelli, A. Sassaroli, S. Del Bianco, Y. Yamada, and G. Zaccanti.
\newblock \textit{Solution of the time-dependent diffusion equation for layered diffusive media by the eigenfunction method.}
\newblock Physical Review E \textbf{67}(5), 056606 (2003).

\bibitem{Carr2016}
E. J. Carr and I. W. Turner.
\newblock \textit{A semi-analytical solution for multilayered diffusion with a moving boundary.}
\newblock International Journal of Heat and Mass Transfer \textbf{102}, 1169--1179 (2016).

\bibitem{Egolf2018}
P. Egolf and K. Hutter.
\newblock \textit{Tsallis extended thermodynamics applied to 2-d turbulence: L\'evy statistics and q-fractional generalized Kraichnanian energy and enstrophy spectra.}
\newblock Entropy \textbf{20}(2), 109 (2018).

\bibitem{Redner2001}
S. Redner.
\newblock \textit{A Guide to First-Passage Processes.}
\newblock Cambridge University Press (2001).

\bibitem{Stantec2022}
J. C. Crittenden, R. R. Trussell, D. W. Hand, K. J. Howe, G. Tchobanoglous, B. Ward, and J. H. Borchardt.
\newblock \textit{Stantec's Water Treatment: Principles and Design.}
\newblock Wiley (2022).

\bibitem{Hurtado2015}
F. J. Hurtado, A. S. Kaiser, and B. Zamora.
\newblock \textit{Fluid dynamic analysis of a continuous stirred tank reactor for technical optimization of wastewater digestion.}
\newblock Water Research \textbf{71}, 282--293 (2015).

\bibitem{Szollosi1982}
A. Sz\"{o}ll\"{o}si-Nagy.
\newblock \textit{The discretization of the continuous linear cascade by means of state space analysis.}
\newblock Journal of Hydrology \textbf{58}, 223--236 (1982).

\bibitem{Szilagyi2010}
J. Szilagyi and A. Sz\"{o}ll\"{o}si-Nagy.
\newblock \textit{Recursive Streamflow Forecasting: A State-Space Approach.}
\newblock Taylor \& Francis (2010).

\end{thebibliography}
\end{document}